\documentclass[twocolumn]{article}
\usepackage{graphicx} 

\title{A Higher Spin-Statistics Theorem for Invertible Quantum Field Theories}
\author{}
\date{\today}
\usepackage{lipsum}\usepackage{tikz}
\usetikzlibrary{matrix,arrows,decorations.pathmorphing,shapes.geometric}

\usepackage{etex}
\usepackage{subcaption}
\usepackage[utf8]{inputenc}
\usepackage{a4wide}
\usepackage{graphicx}
\usepackage{wrapfig}
\usepackage[hmarginratio={1:1},     
vmarginratio={1:1},     
textwidth=17.6cm,        
textheight=22cm,
heightrounded,]{geometry}

\usepackage{amsmath,accents}
\usepackage{amsthm}
\usepackage{amssymb}

\usepackage{mathrsfs, mathtools}
\usepackage{overpic}

\usepackage[utf8]{inputenc}

\usepackage[all,cmtip]{xy}

\makeatletter
\DeclareRobustCommand{\em}{%
	\@nomath\em \if b\expandafter\@car\f@series\@nil
	\normalfont \else \slshape \fi}
\makeatother

\usepackage[hidelinks]{hyperref}

\numberwithin{equation}{section}
\usepackage{mathtools}
\mathtoolsset{showonlyrefs}
\numberwithin{equation}{section}

\newtheoremstyle{style1}
{13pt}
{13pt}
{}
{}
{\normalfont\bfseries}
{.}
{.5em}
{}

\theoremstyle{style1}

\newtheorem{definition}{Definition}[section]

\newtheorem{remark}[definition]{Remark}

\makeatletter
\newtheorem*{repd@theorem}{\repd@title}
\newcommand{\newrepdtheorem}[2]{%
	\newenvironment{repd#1}[1]{%
		\def\repd@title{#2 \ref{##1}}%
		\begin{repd@theorem}}%
		{\end{repd@theorem}}}
\makeatother

\newrepdtheorem{definition}{Definition}

\newcommand{\catf}[1]{{\mathsf{#1}}}

\newtheoremstyle{style2}
{13pt}
{13pt}
{\slshape}
{}
{\normalfont\bfseries}
{.}
{.5em}
{}

\theoremstyle{style2}

\makeatletter
\newtheorem*{rep@theorem}{\rep@title}
\newcommand{\newreptheorem}[2]{%
	\newenvironment{rep#1}[1]{%
		\def\rep@title{#2 \ref{##1}}%
		\begin{rep@theorem}}%
		{\end{rep@theorem}}}
\makeatother

\newreptheorem{theorem}{Theorem}
\newreptheorem{corollary}{Corollary}

\newtheorem{theorem}[definition]{Theorem}

\usepackage{tikz}
\usetikzlibrary{matrix,arrows,decorations.pathmorphing,shapes.geometric}
\usepackage{tikz-cd}

\usepackage{enumitem}

\usepackage{needspace}

\newcommand{\R}{\mathbb{R}}
\newcommand{\C}{\mathbb{C}}
\newcommand{\Z}{\mathbb{Z}}

\newcommand{\Or}{\mathrm{O}}

\newcommand{\colim}{\operatorname{colim}}

\usepackage{pifont}

\newcommand{\id}{\operatorname{id}}

\let\to\undefined
\newcommand{\to}{\longrightarrow}

\let\colon\undefined\newcommand{\colon}{:}

\DeclareMathSymbol{\Phiit}{\mathalpha}{letters}{"08} 
\DeclareMathSymbol{\Psiit}{\mathalpha}{letters}{"09}
\DeclareMathSymbol{\Sigmait}{\mathalpha}{letters}{"06}
\DeclareMathSymbol{\Xiit}{\mathalpha}{letters}{"04}
\DeclareMathSymbol{\Piit}{\mathalpha}{letters}{"05}\let\Pi\undefined\newcommand{\Pi}{\Piit}
\DeclareMathSymbol{\Gammait}{\mathalpha}{letters}{"00}
\DeclareMathSymbol{\Omegait}{\mathalpha}{letters}{"0A}\let\Omega\undefined\newcommand{\Omega}{\Omegait}
\DeclareMathSymbol{\Upsilonit}{\mathalpha}{letters}{"07}
\DeclareMathSymbol{\Thetait}{\mathalpha}{letters}{"02}
\DeclareMathSymbol{\Lambdait}{\mathalpha}{letters}{"03}\let\Lambda\undefined\newcommand{\Lambda}{\Lambdait}

\let\Phi\undefined\newcommand{\Phi}{\Phiit}
\let\Sigma\undefined\newcommand{\Sigma}{\Sigmait}
\let\Psi\undefined\newcommand{\Psi}{\Psiit}
\let\Gamma\undefined\newcommand{\Gamma}{\Gammait}

\newcommand{\Spin}{\operatorname{Spin}}

\newcommand{\Sp}{\mathbf{Sp}}
\newcommand{\fib}{\operatorname{fib}}
\newcommand{\Bord}{\mathbf{Bord}}

\usepackage{multicol}

\makeatletter
\renewcommand\section{\@startsection {section}{1}{\z@}%
	{-3.5ex \@plus -1ex \@minus -.2ex}%
	{2.3ex \@plus.2ex}%
	{\normalfont\scshape\centering}}
\makeatother

\usepackage{titletoc}

\dottedcontents{section}[1em]{\rmfamily}{1em}{0.2cm}
\dottedcontents{subsection}[0em]{}{3.3em}{1pc}

\usepackage{titlesec}
\titleformat{\subsection}[runin]
{\normalfont\bfseries}
{\thesubsection}
{0.5em}
{}
[.]

\definecolor{Blue}  {rgb} {0.282352,0.239215,0.803921}
\definecolor{Green} {rgb} {0.133333,0.545098,0.133333}
\definecolor{Red}   {rgb} {0.803921,0.000000,0.000000}
\definecolor{Violet}{rgb} {0.580392,0.000000,0.827450}

\usepackage{multicol}

\usepackage{titletoc}
\dottedcontents{section}[1em]{\rmfamily}{1.5em}{0.2cm}
\dottedcontents{subsection}[5em]{\rmfamily}{1.9em}{0.2cm}

\author{Cameron Krulewski, Lukas M\"{u}ller, Luuk Stehouwer}

\begin{document}
	
\twocolumn[
  \begin{@twocolumnfalse}
	\maketitle
 \begin{abstract}
    \noindent We prove that every unitary invertible quantum field theory satisfies a generalization of the famous spin-statistics theorem. To formulate this extension, we define a \emph{higher spin} action of the stable orthogonal group $O$ on appropriate spacetime manifolds, which extends both the reflection involution and spin flip. On the algebraic side, we define a \emph{higher statistics} action of $O$ on the universal target for invertible field theories, $I\Z$, which extends both complex conjugation and fermion parity $(-1)^F$. 
    We prove that every unitary invertible quantum field theory intertwines these actions. 
 \end{abstract}
  \end{@twocolumnfalse}
]

	\section{Introduction}
	Particles come in two types,\footnote{Excluding anyons in $2+1$d.} distinguished by their exchange statistics (braiding):
    bosons satisfy $\phi \phi'=\phi' \phi$, while fermions satisfy $\psi \psi'=-\psi' \psi$.
	Mathematically, this distinction is encoded using super vector spaces $V=V_B\oplus V_F$ as state spaces, where the elements of {the even piece} $V_B$ are bosonic and those of {the odd piece} $V_F$ are fermionic.
	The eigenvalues of the fermion parity operator $(-1)^F$ are $1$ on $V_B$ and $-1$ on $V_F$,
    detecting
 the exchange statistics
    of particles. 
    This operator is part of a $\Z_2^F$ 1-form symmetry (an action of the 2-group $B\Z_2^F$) on the category $\catf{sVect}$ of super vector spaces~\cite{deligne2000quantum}. 
 
 Particles on $\R^n$ transform under the 
 special orthogonal group\footnote{Throughout this paper we work in Euclidean signature.} $\operatorname{SO}_n$, or more precisely under the spin group, its universal cover (for $n\geq 3$)
 $$\Z_2 \rightarrow \operatorname{Spin}_n\rightarrow \operatorname{SO}_n. $$ 
 The non-contractible loop in $\operatorname{SO}_n$ corresponding to a 360-degree rotation lifts to a path in $\Spin_n$ 
 {whose endpoint we call} $c\in \Z_2$. 
	The action of $c$ hence encodes the transformation of particles under `360-degree rotations': particles have \emph{even spin} if $c$ acts trivially and \emph{odd spin} if $c$ acts by multiplication with $-1$.   
	For quantum field theories on curved spin manifolds, the orthogonal group does not act globally.
 However, the element $c$ still acts by an automorphism of the spin structure, which assembles into an action of the 2-group $B \Z_2^c$.  

The spin-statistics theorem relates these a priori unrelated $B\Z_2$ actions:
a quantum field theory $\mathcal{Z}$ associates to a time-slice $\Sigma$ a super-Hilbert space $\mathcal{Z}(\Sigma)$ on which $c$ acts by an automorphism $\mathcal{Z}(c)$. In every unitary quantum field theory this action agrees with the action of $(-1)^F$; i.e.\ the particle's transformation under 360-degree rotations determines its statistics. Mathematically, this means that every unitary quantum field theory $\mathcal{Z}$ intertwines the two 1-form symmetries.
	
Unitary quantum field theories also intertwine a $\Z_2$ 0-form symmetry: the geometric operation of orientation reversal corresponds to the algebraic operation of complex conjugation. 
These two $\Z_2$ groups correspond to the first two of the (8-periodic) stable homotopy groups
\begin{align*}
    \pi_0(\Or)&= \Z_2, \  \pi_1(\Or)= \Z_2, \ \pi_2(\Or)= 0, 
    \\
    \pi_3(\Or)&= \Z, \  \pi_4(\Or)= 0,\ \pi_5(\Or)= 0,  
    \\
    \pi_6(\Or)&= 0,  \ \pi_7(\Or)= \Z , \ \pi_8(\Or)= \Z_2, \, \dots 
\end{align*} 
of the stable orthogonal group $\Or$.
We can thus express the combined facts that orientation reversal corresponds to complex conjugation and 360-degree rotation corresponds to fermion parity as equivariance for the truncation $\pi_{\leq 1}(\Or)$. 
The fact that this equivariance holds for unitary (non-extended) topological field theories was shown in \cite{luukspinstatistiscs}. The special case of {equivariance for} invertible theories was previously shown in \cite[Section 11]{freed_reflection_2021}. 

In this paper, we extend the latter spin-statistics theorem from $\pi_{\leq 1} (\Or)$ to all of $\Or$ and to fully-extended field theories:\footnote{A connection between $\Or$ and spin statistics was conjectured in~\cite{theospinstatistics}. See also  \cite[Section 3.3]{kapranov2021supergeometry}.} we establish a geometric action of $\Or$ on 
manifolds with stable tangential structures (including but not limited to spin structures) and show that every unitary invertible field theory intertwines this with an algebraic $\Or$-action on extended operators.

This latter $\Or$-action generalizing complex conjugation and fermion parity is not well understood for general theories.
This is the main reason for our restriction to invertible field theories, which can be accessed via stable homotopy theory and which are of separate interest because of their relationship with anomalies and symmetry-protected topological phases \cite{freed_short-range_2014}.
We expect our results to generalize to non-invertible theories. 

In more detail: the universal target for invertible $n$-dimensional theories is the spectrum $\Sigma^{n+1}I\Z$, (a suspension of) the Anderson dual of the sphere.
The homotopy groups\footnote{Which are concentrated in degrees $\leq 1$.} $\pi_{-k}$ of $\Sigma^{}I\Z$ encode the types of algebraic structures an invertible quantum field theory can assign to submanifolds of codimension $k$ in spacetime. 
They are closely related to the stable homotopy groups of spheres:
\begin{align*}
    &\pi_1 (\Sigma I\Z) = \Z,~ \pi_{0} (\Sigma I\Z) = 0, ~ \pi_{-1} (\Sigma I\Z) = \Z_2 \\
    &\pi_{-2} (\Sigma I\Z) = \Z_2, ~ \pi_{-3} (\Sigma I\Z) = \Z_{24}, ~ \dots
\end{align*} 
As an alternative to $\Sigma^{n+1}I\Z$, it is also common to work with $\Sigma^n I\C^\times$, the Brown-Comenetz dual of the sphere. 
    Intuitively, $\Sigma^{n+1}I\Z$ is analogous to $\Sigma^n I\C^\times$, except that we take the { continuous} topology on $\C^\times$ for $\Sigma^{n+1}I\Z$.
  We do not know whether our theorem also holds for unitary theories with values in $\Sigma^{n} I\C^\times$, but it does when we restrict to $\Sigma^{n} IU(1)$.\footnote{The target $IU(1)$ satisfies the same universal property as $I\C^\times$ with $U(1)$ replacing $\C^\times$. Note that by restricting to unitary theories with target $\Sigma^n IU(1)$, we exclude the unstable theories that arise for target $I\C^\times$, which fall outside of the scope of our proof~\cite[Theorem 7.22 and 8.29]{freed_reflection_2021}.} 

The $\Z_2$ of $\pi_{-1} (\Sigma I\Z)$ encodes the distinction into bosons and fermions at the level of state spaces (which are assigned to codimension one manifolds). More specifically, the $J$-homomorphism (which we discuss in Section~\ref{sec:Jhom}) defines the $\Or$-action on $\Sigma I\Z$, or alternatively on $IU(1)$, and in the latter case detects the Fermi-Bose distinction through the inclusion map from $\pi_1( \Or) = \Z_2$ to $U(1)$.

Let us extend this interpretation to the next higher non-trivial homotopy group, $\pi_3(\Or)=\Z$, informally. 
An invertible quantum field theory assigns to manifolds of codimension $k$ higher categorical generalizations of super lines: elements of $\pi_{-k} (\Sigma I\Z)$. 
The first interesting new structures are classified by $\pi_{-3}(\Sigma I\Z)=\pi_{-3}( IU(1))=\Z_{24}$. 
We work with $IU(1)$ for simplicity, in which case the higher statistics action by $1\in \pi_3(\Or)= \Z$ is the inclusion $\Z_{24}\to U(1)$ (see Section~\ref{sec:Jhom} for details).    
The higher spin-statistics theorem implies that for every codimension 3 manifold $S$, the value of the higher line associated to $\mathcal{Z}(S)$ {in $\Z_{24}$} is equal to the evaluation of $\mathcal{Z}$ on the higher spin action of $1\in \pi_3(\Or)$, which is a 3-automorphism of $S$. In many cases, for example for theories depending \emph{only} on spin structures, we can ensure that the higher spin action is trivial on $\pi_3$ and hence the higher line $\mathcal{Z}(S)$ is forced to be trivial. 

\vspace*{0.2cm}\textsc{Acknowledgments.}
We thank Theo Johnson-Freyd for sharing his ideas related to spin-statistics with us, offering encouragement, and providing valuable feedback.  
We thank Daniel Brennan and Andrea Grigoletto  
        for helpful discussions related to this project and Yu Leon Liu for comments on the first draft.
	 	The authors gratefully acknowledge support of the Simons Collaboration on Global Categorical Symmetries and of the Perimeter Institute. Research at the Perimeter Institute is supported in part by the Government of Canada through the Department of Innovation, Science and Economic Development and by the Province of Ontario through the Ministry of Colleges and Universities. The Perimeter Institute is in the Haldimand Tract, land promised to the Six Nations. 
        During the writing of this paper, CK was supported by the National Science Foundation GRFP under grant number DGE-2141064.
        LS is supported by the Atlantic Association for Research in the Mathematical Sciences. 
LS is also grateful to Dalhousie University for providing the facilities to carry out his research.

\section{Setup}
\subsection{Higher spin} 
 \label{sec:O-action}
 Quantum field theories depend on tangential structures, such as spin structures or orientations. In dimension $n$ these are encoded by a topological group $H_n$ and a continuous group homomorphism $H_n\to \Or_n$ (an $n$-dimensional representation). A \emph{(topological) $H_n$-structure} on an $n$-dimensional manifold $M$ is a principal $H_n$-bundle $P_M$ over $M$ together with a vector bundle isomorphism $P_M\times_{H_n}\R^n \xrightarrow{\cong} TM$. This description has a convenient reformulation in terms of homotopy theory: an $H_n$-structure is equivalently a homotopy lift of the map $M\to B\Or_n$ classifying the tangent bundle along the map $BH_n \to B\Or_n$. 

 Stable tangential structures, such as orientations, spin structures, or pin structures, come in families indexed by $n$. They are described in the $n\to \infty$ limit by a map of spaces $\rho\colon BH\to B\Or$, where $\Or$ is the stable orthogonal group. The finite $n$-version can be recovered as the (homotopy) pullback 
 \begin{equation}
    \begin{tikzcd}
        B{H}_n \ar[d,"\rho_n",swap] \ar[r]&  BH \ar[d,"\rho"] \\ 
        B\Or_n  \ar[r] & B\Or .
    \end{tikzcd} \ \ 
\end{equation}
 
 To define $\Or$-actions on manifolds with stable tangential structures, we use that
 the assignment sending a space $BH_n$ over $B\Or_n$ to the collection of $n$-dimensional manifolds with that $H_n$-tangential structure (or the spectrum $MTH_n$, as in Section~\ref{sec:invertibleTFT}) is functorial in $\rho_n\colon BH_n\to B\Or_n$. Hence we can specify an action of the stable orthogonal group $\Or$ on manifolds with $H_n$-structure by acting on $\rho_n$. 

A homotopy-theoretic way to define actions of a topological group $G$ 
on some object $X$ is to provide a fibration with base $BG$ and fiber $X$.
In the setting of spaces over $B\Or_n$, the following formulation is convenient. 
\begin{definition}
    An \emph{action of $G$ on $\rho_n\colon BH_n \to B\Or_n$} is a fibration
 $X' \xrightarrow{(\rho',\alpha)} B\Or_n \times BG$
fitting into a pullback square 
\begin{equation}
\label{eq:action}
    \begin{tikzcd}
        B{H}_n \ar[d,"\rho_n"] \ar[r]&  X' \ar[d, "{(\rho',\alpha)}"] \\ 
        B\Or_n  \ar[r,"i_1"] & B\Or_n \times BG
    \end{tikzcd}
\end{equation}
where $i_1$ is the inclusion in the first factor using the basepoint $* \in BG$.
\end{definition}

We now define a specific action of $\Or$ on an arbitrary stable tangential structure $BH \to B\Or$, which we call the \emph{higher spin action} and denote by $\beta$.
This action extends the action of $\Or_1$ defined in \cite[Appendix E]{freed_reflection_2021} and used in \cite[Equation (8)]{ferrer_dagger_2024}.
{It also recovers the usual spin flip action of $\pi_1(\Or)$; see \cite[Example 3.3]{theospinstatistics} and the explanation below for details.}
To define the action, we define $B\tilde{H}_n$ as the pullback 
\begin{equation}
    \begin{tikzcd}
        B\tilde{H}_n \ar[d] \ar[r]&  B{H} \ar[d,"\rho"] \\ 
        B\Or_n \times B\Or \ar[r, "-\ominus -"] & B\Or.
    \end{tikzcd}
\end{equation}  
Note that in the stable limit this pullback is just $B\tilde{H} = BH\times B\Or \xrightarrow{\tiny\begin{pmatrix}
   \rho & \id_{B\Or} \\
   0 & \id_{B\Or}
\end{pmatrix}\normalsize} 
B\Or\times B\Or$.

This indeed defines an action because the further pullback {along the inclusion $i_1$}
\begin{equation}
    \begin{tikzcd}
        BH_n \ar[r] \ar[d] & B\tilde{H}_n \ar[d] \ar[r]&  B{H} \ar[d,"\rho"] \\ 
        B\Or_n \ar[r,"i_1"] & B\Or_n \times B\Or \ar[r, "-\ominus -"] & B\Or
    \end{tikzcd}
\end{equation}  
is $\rho_n\colon BH_n \to B\Or_n$ by composition of pullbacks.

	\subsection{Invertible field theories via stable homotopy theory} 
 \label{sec:invertibleTFT}
	
In general terms, a field theory is invertible if it admits an inverse under the stacking operation.
In \cite{freed_reflection_2021}, it is argued that the data of an invertible field theory is uniquely specified by its partition function.
More specifically, if we study a theory with $H_n$-background gauge fields for some tangential structure $\rho_n\colon BH_n \to B\Or_n$, the partition function is a continuous map of commutative monoids from the commutative monoid of closed $n$-dimensional manifolds with $H_n$-structure to $\C^\times$ satisfying a cut-and-paste relation. 
The authors in \cite[\S5.3]{freed_reflection_2021} show that this can be rephrased using the language of algebraic topology as a map of spectra
\[
\Sigma^n 
MTH_n \to \Sigma^{n+1} I\Z,
\]
where the domain spectrum is the Madsen-Tillmann spectrum of $H_n \to \Or_n$~\cite{galatius_homotopy_2009}. 
This makes invertible theories amenable to powerful computational methods in stable homotopy theory. 
If $BH_n \to B\Or_n$ is induced by a stable tangential structure $BH \to B\Or$, these spectra assemble into a tower of which we often want to take the stable colimit $MTH \coloneqq \colim_n \Sigma^n MTH_n$. The main theorem of~\cite{freed_reflection_2021} relates this colimit to unitarity.\footnote{Their results apply strictly speaking to only a subset of stable tangential structures. However, their setup and proof generalizes straightforwardly.}

\begin{theorem}[{\cite[Theorem 8.20]{freed_reflection_2021}}]\label{FH820}
    An invertible field theory
    \[
     \Sigma^n MTH_n \to \Sigma^{n+1} I\Z
    \]
    is unitary\footnote{Freed--Hopkins phrase this theorem for reflection positive theories, which are the Wick-rotated analog of unitary theories. However, we make no distinction between the terms \emph{unitary} and \emph{reflection positive} in this article.}
    if and only if it factors through $MTH$.
\end{theorem}
This theorem shows that by restricting to the more physically-realistic setting of unitary theories, we can classify invertible field theories by bordism groups (the homotopy groups of the spectrum $MTH$). 
In contrast, non-unitary theories are classified by SKK or Reinhart bordism groups \cite{kst, freedsbook, MR153021, luukreneesimona} instead.

\subsection{The \texorpdfstring{$J$}{J}-homomorphism and higher statistics}
\label{sec:Jhom}

Elements of the orthogonal group induce homeomorphisms of spheres.
In the stable setting, these induce a map
$$ B\Or \to BGL_1(\mathbb{S}) $$ called the \emph{$J$-homomorphism}, where $GL_1(\mathbb{S})$ denotes the automorphisms of the sphere spectrum~\cite{whiteheadjhom, mathewjhom}. 
By mapping these automorphisms to the $\infty$-category of all spectra we obtain a map
\begin{align}
 J\colon B\Or \to \Sp
 \label{eq: J}
\end{align}
that sends the base point of $B\Or$ to the sphere spectrum $\mathbb{S}$. 
The functor $J$ sends direct sums $\oplus$ of vector bundles to tensor products $\wedge$ of spectra.
Given a spectrum $E \in \Sp$, let $J_E\colon B\Or \to \Sp$ be the composition of $J$ with the tensoring functor $(-) \wedge E\colon\Sp \to \Sp$. 
This defines an action of $\Or$ on any spectrum $E$ that is functorial in the spectrum $E$.
Note that $J_\mathbb{S}$ recovers the original map $J$ in Equation~\eqref{eq: J}. 
Importantly, any map of spectra $f\colon E\to E'$ intertwines the actions $J_E$ and $J_{E'}$; i.e.\ $f$ has a canonical structure of an $\Or$-equivariant map.  

The $J$-homomorphism allows for an elegant definition for the Madsen-Tillmann spectrum of a tangential structure $\rho_n \colon BH_n \to B\Or_n$: the spectrum $\Sigma^n MTH_n$
is the colimit of $\ominus J \circ \rho_n$~\cite{lewis_equivariant_1986, ando_infty-categorical_2014,ando_units_2014}. {Here and below $\ominus J$ denotes the precomposition of $J$ with the automorphism $\ominus \colon BO \to BO$ that takes the `orthogonal complement,' reflecting the fact that the Madsen-Tillmann spectrum is a tangential version of the normal Thom spectrum.}
This description makes the actions constructed in Section~\ref{sec:O-action} manifest. 
If $(\rho',\alpha)\colon B\tilde{H}_n \to  B\Or_n \times BG $ is an action of $G$ on $\rho_n$, there is an induced $G$-action $BG \to \Sp$ on $\Sigma^n MTH_n$.
The corresponding functor $BG \to \Sp$ sending the base point to $\Sigma^n MTH_n$ is defined by left Kan extension of $\ominus J \circ \rho'$ along $\alpha$:
\begin{equation}
\begin{tikzcd}[column sep = 35]
    BH_n \ar[d] \ar[rd,"\ominus J \circ \rho", bend left] & 
    \\
    B \tilde{H}_n \ar[r,"\ominus J \circ \rho'"] \ar[d,"\alpha"] & \Sp
    \\
    BG \ar[ru,dashed, "\operatorname{Lan}_{\alpha} \ominus J \circ \rho'", swap, bend right] &
\end{tikzcd}
\label{Eq: Kan extension}
\end{equation}
Now specialize to $G=\Or$. We define the \emph{higher statistics $\Or$-action} on $\Sigma I\Z$ by $\ominus  J_{\Sigma  I\Z}$. Similarly, we define the higher statistics action on $IU(1)$ as $\ominus  J_{ IU(1)}$. We will comment on the interpretation of these actions in the next section. 

To work out the action of $\ominus  J_{ IU(1)}$ on $IU(1)$, we use that $IU(1)$ is characterized by the universal property $\pi_{-k} \operatorname{Map}(X,IU(1)) \cong \operatorname{Hom}(\pi_k(X), U(1) )$ for all spectra $X$. The action $\ominus  J_{ IU(1)}$ on this space by postcomposition can be identified with the action by precomposition with $\ominus  J_{X}$, since $J$ commutes with all maps between spectra. Setting $X=\mathbb{S}$, we conclude that the action of $\ominus J_{IU(1)}$ on the homotopy groups of $IU(1)$ is given by pullback along the action on $\mathbb{S}$. {Adams~\cite[Theorem 1.5]{adamsjhom} computed the order of the image of the $J$-homomorphism on homotopy groups, showing that $\Z = \pi_3(\Or) \to \pi_3(\mathbb{S}) = \Z_{24}$ is surjective.} It now follows that the map $\pi_{-3}(IU(1)) = \Z_{24}\to U(1) = \pi_0(IU(1))$ corresponding to the action with a generator of $\pi_3(\Or)$ is an inclusion. 

\subsection{Definition of higher spin-statistics}

In this section, we define higher spin-statistics and explain its relation to previous definitions.
\begin{definition}
A \emph{higher spin-statistics connection} for an invertible field theory $\mathcal{Z}\colon\Sigma^n MTH_n \to \Sigma^{n+1} I\Z$ is equivariance data for the higher spin action $\beta$ on $\Sigma^n MTH_n$ and higher statistics action $\ominus J_{\Sigma^{n+1}I\Z}$ on $\Sigma^{n+1} I\Z$. 
\end{definition}
\begin{remark}
Note that a higher spin-statistics connection is a structure on (instead of a property of) an invertible field theory; in general it is possible to equip the same theory with different higher spin-statistics connections.  
\end{remark}

As mentioned in the introduction, restricting to $\Or_1=\Z_2 \subset \Or $ recovers the relation on field theories between orientation reversal of spacetime and complex conjugation.
In its most elementary form, it says that partition function $\mathcal{Z}$ should be $\pi_0(\Or) = \Z_2$-equivariant: 
\[
\mathcal{Z}(\overline{M}) = \overline{\mathcal{Z}(M)},
\]
where $\overline{M}$ is the restriction of the higher spin action to $\Or_0=\Z_2$, which agrees with the orientation reversal defined in~\cite[Section 4.1]{freed_reflection_2021}. That the action of $\ominus J_{I\Z}$ on $I\Z$ corresponds to complex conjugation is explained in detail in~\cite[Section 6.3]{freed_reflection_2021}.

In order to describe physical aspects of further $\Or$-equivariance data, we rephrase the definition of Section \ref{sec:invertibleTFT} in the functorial field theory formalism. 
Let $\Bord^H_n$ denote the fully local bordism $n$-category with $H$-background gauge fields for some stable tangential structure $\rho\colon  BH \to B\Or$~\cite{lurie_classification_2009,calaque_note_2019}. A functorial quantum field theory (with target a symmetric monoidal $(\infty,n)$-category $\mathcal{T}$) is a symmetric monoidal functor $\mathcal{Z} \colon \Bord_n^H \to \mathcal{T}$.
The connected cover of $\Sigma^{n+1} I\Z$ seen as a symmetric monoidal $(\infty,n)$-category is the universal target for invertible quantum field theories.
A consequence
of the theorem that the localization $\|\Bord^H_n\|$ of the bordism category $\Bord^H_n$ is the Madsen-Tillmann spectrum $\Sigma^n MTH_n$ \cite{galatius_homotopy_2009,schommerpriesinvertible} is that invertible quantum field theories are the same as maps of spectra $\Sigma^n MTH_n \to \Sigma^{n+1} I\Z$.
The advantage of the functorial description is that it allows for the physical interpretation of the other homotopy groups of $I\Z$ in terms of higher-codimension submanifolds of spacetime.
For example, the coconnective truncation of $\Sigma^2 I\Z$ is equivalent to the Picard groupoid of one-dimensional super vector spaces, with the continuous topology on morphism spaces.
Therefore, given an $n$-dimensional {partition function satisfying cut and paste relations}, we obtain the surprising result that for every $(n-1)$-dimensional closed manifold $Y$, there is a unique parity of the one-dimensional state space $\mathcal{Z}(Y)$ that defines consistent TFT.

The topological Picard groupoid of super lines has a canonical $\Z_2\times B\Z_2$-action by complex conjugation and fermion parity, which agrees with the action induced by $\ominus J$ on the connective cover of $\Sigma^2 I\Z$.
Hence a higher spin-statistics connection\color{black}, when restricted to $B\Z_2$ and state spaces, gives the isomorphism $\mathcal{Z}(\beta(-1_{B\Z_2})_Y) \cong \ominus J(-1_{B\Z_2})_{\mathcal{Z}(Y)} =  (-1)^F_{\mathcal{Z}(Y)}$.   
For fermionic structure groups~\cite[Section 3]{muller_reflection_2023} like $ H=\Spin$, the action of $\beta (-1_{B\Z_2})$ is given by acting by the central element $c\in H$. This allows us to conclude the {usual} spin-statistics connection: $\mathcal{Z}(c) = (-1)^F_{\mathcal{Z}(Y)}$.

\section{Proof of the higher spin-statistics theorem}
Equipped with the material of the previous section, we can now give a precise formulation and proof of the higher spin-statistics theorem. 

\begin{theorem}
    Every unitary \color{black} invertible field theory
    \[
    \mathcal{Z}\colon \Sigma^n MTH_n \to \Sigma^{n+1} I\Z
    \]
     has a canonical higher spin-statistics connection. \color{black}
\end{theorem}

{Here by unitary we mean reflection positive in the sense of \cite{freed_reflection_2021}. In our proof, we do not require the explicit definition, only the result of Freed and Hopkins that such a theory factors through $\Sigma^n MTH_n \to MTH$ up to homotopy. However, to give readers an idea of the nontriviality of our theorem, reflection positivity consists of equivariance as well as positivity data, but only for the subgroup $\Or_1 \subseteq \Or$. The $\Or_1$-equivariance data makes the higher lines into higher Hermitian vector spaces, while the positivity data ensures these are Hilbert spaces. }
Our theorem demonstrates that unitarity theories satisfy equivariance conditions for the entirety of $\Or$, extending the $\Or_1$-equivariance that is part of the hypothesis.

The rest of this section is concerned with the proof of this theorem, based on the work of Freed-Hopkins. First we observe that by the main result (Theorem \ref{FH820}) of~\cite{freed_reflection_2021}, $\mathcal{Z}$ factors through $MTH$ up homotopy.

Since the diagram
\begin{equation}
\begin{tikzcd}[column sep = 0.1]
  B {H}_n \ar[d] \ar[rr]& &  B {H} \ar[d]     \\ 
 B \tilde{H}_n \ar[rr] \ar[rd]  & &  B \tilde{H} \ar[ld]   \\ 
   & B\Or & 
\end{tikzcd}
\end{equation}
commutes, the functoriality of Kan extensions implies that the map of spectra $\Sigma^n MTH_n \to MTH$ is $\Or$-equivariant. 

The action of $\Or$ on $MTH$ defined in Section \ref{sec:O-action} induces the $\Or$-action $\ominus J_{MTH}$, as we can see by computing with the description in~\eqref{Eq: Kan extension}. For $V\in B\Or$,
    \begin{align*}
    \operatorname{Lan}_\alpha &(\ominus J \circ \rho')(V) 
    \\
    &\cong  \colim_{(V,h) \in \fib(\alpha)_V} \ominus J \circ \rho' (V,h) 
    \\
    &\cong \colim_{(V,h) \in \fib(\alpha)_V} \ominus J (\rho (h) \oplus V) 
    \\
    &\cong \colim_{(V,h) \in \operatorname{fib}(\alpha)_V} \ominus J (\rho (h)) \wedge \ominus J(V) 
    \\
    &\cong (\colim_{h \in BH} \ominus J(\rho(h))) \wedge \ominus J(V) 
    \\
    &\cong MTH \wedge \ominus J(V) = \ominus J_{MTH}(V).
    \end{align*}
Here we used that tensoring with any spectrum preserves colimits.

As explained in Section \ref{sec:Jhom}, every map of spectra is equivariant for the action of the $J$-homomorphism and hence also for $\ominus J$, which allows us to conclude the higher spin-statistics theorem from the factorization above.  

Let us stress that the $\Or$-action on $\Sigma^n MTH_n$ is not equivalent to that of $\ominus J_{\Sigma^n MTH_n}$. If it were, that would imply that every invertible field theory  would have a canonical higher spin-statistics connection. 
It is easy to construct examples of (non-unitary) invertible field theories that do not even satisfy higher spin-statistics on $\pi_0(\Or)$ and $\pi_1(\Or)$; see e.g. \cite[Example 11.2]{freed_reflection_2021}.

\section{Consequences and outlook}

A consequence of the spin-statistics theorem is that a unitary theory 
without a dependence on
(possibly twisted) spin structures\footnote{For example, these include pin$^+$ and pin$^-$ structures.}
cannot involve fermions. This is because without a (twisted) spin structure dependence, the generator $c\in \pi_1(\Or)$ acts trivially on the geometric side, while $\ominus J_{I\Z}$ acts nontrivially on the state spaces (since it acts by
$-1$ on odd super vector spaces). 

Our theorem allows us to generalize this argument: recall that $\pi_3(\Or)=\Z$, which can be used to distinguish 
the $\Z_{24}$ different elements of $\pi_{-3}(\Sigma I\Z)$. 
If the geometric action of this $\Z$ on $\Sigma^n MTH_n$ is trivial, then 
excitations with this kind of higher statistics
are prohibited. This is for example the case if the stable tangential structure on spacetime is built from an internal symmetry group that includes only fermion parity and time-reversing elements.\footnote{This kind of tangential structure is of interest because it includes for example the tenfold way classes for fermionic topological insulators and superconductors.}
A convenient way of encoding this kind of tangential structure is through a fermionic group $G$~\cite[Section 3]{muller_reflection_2023}, which homotopically is equivalent to a group $G_b$ together with a map $BG_b\to B\Z_2\times B^2\Z_2= \pi_{\leq 2}(B\Or)$. The corresponding stable tangential structure is defined as the pullback 
\begin{equation}
\begin{tikzcd}
    BH \ar[d]\ar[r] & BG_b \ar[d] \\ 
    B\Or \ar[r] & \pi_{\leq 2}(B\Or)
\end{tikzcd}
\end{equation}
Then, the group defining the action on $\Sigma^n MTH_n$ sits in a pullback diagram 
\begin{equation}
\begin{tikzcd}
    B\tilde{H}_n \ar[d]\ar[r] & X \ar[r] \ar[d] & BG_b \ar[d] \\ 
    B\Or_n\times B\Or \ar[r] & B\Or_n\times \pi_{\leq 2}(B\Or) \ar[r, "-\ominus-"]  & \pi_{\leq 2}(B\Or)
\end{tikzcd}
\end{equation} 
and thus the $\Or$-action factors through an action of its truncation $\pi_{\leq 1}(\Or)$.
Therefore, in all theories with this kind of tangential structure,
we cannot see any of these higher versions of fermions.
Instead, to see these higher structures, we must incorporate information from higher pieces of the Whitehead tower of the orthogonal group, which at the third stage means that spacetime should carry (twisted) string structures.
A way of formulating this intuitively is that 
{just as consistent theories of fermions should be formulated on manifolds with twisted spin structures, theories of higher fermions should be formulated on manifolds with twisted string structures.}
{See e.g.\ \cite{debray2024smith} and \cite[Section 7]{kapustin2015fermionic} for more on twisted spin structures in this context, and see e.g.\ \cite{MR2966944} and \cite[Remark 2.16]{anotherthingofArunandMatt} 
for the generalization to twisted string structures.}

Since the $J$-homomorphism $\Z \cong \pi_3(O) \to \pi_3(\mathbb{S}) \cong \Z_{24}$ is surjective, string-statistics can be described concretely as follows.
The Pontryagin-Thom theorem implies that the $n$th stable homotopy group of spheres is isomorphic to the bordism group of $n$-dimensional framed manifolds.
The framed three-dimensional bordism group $\Omega^{\textnormal{fr}}_3 \cong \Z_{24}$ has an explicit generator given by the manifold $S^3 = SU(2)$ with the framing induced by the Lie group structure. 
Therefore, string statistics states that the `$3$-vector space' associated to a codimension 3 manifold $M$ can be computed as follows.
We first perform the dimensional reduction procedure of taking the product with $S^3$ with the tangential structure induced by the Lie group framing to get a QFT in three dimensions lower.
We then evaluate the partition function of this QFT on $M$, which turns out to necessarily result in a $24$th root of unity.

This procedure should be compared with the following procedure, which allows for the computation of the parity of the state space on a codimension one manifold $Y$.
The framed one-dimensional bordism group $\Omega^{\textnormal{fr}}_1 \cong \Z_2$ is generated by $S^1 = U(1)$ with the (Ramond or periodic) Lie group framing.
Then compactifying on $S^1$
\[
Z(Y \times S^1) = \operatorname{str}(\id_{Z(Y)})
\]
computes the parity of the line $Z(Y)$. Here $\operatorname{str}$ indicates the supertrace.

We expect that our result extends to non-invertible quantum field theories, but such an extension seems far out of reach of current mathematics. Even in the setting of topological field theories, we lack a good target category equipped with a higher statistics action.  
In the past few years there has been significant work in extending the universal target to the non-invertible case \cite{Teleman_2022,Johnson-Freyd_2023}. 
This target is supposed to be the higher categorical algebraic closure of the real numbers, while the higher statistics action is expected to be part of the Galois action. In low dimensions this was verified in~\cite{theospinstatistics}.     
We are unaware of any results taking the topology of $\C^\times$ into account. 

A more detailed physical interpretation of our result is a wide open problem, especially for non-invertible quantum field theories. A first step might be to understand the role of twisted string structures in the action of $\pi_3 (\Or)$ in examples.

\subsubsection*{Data availability}

This work does not generate any datasets. One
can obtain the relevant materials from the references below.

\subsubsection*{Conflict of interest}

The authors have no competing interests to declare that are relevant to the content of this article.

\footnotesize
\bibliographystyle{alpha}
\bibliography{bib}

\end{document}